\documentstyle[12pt]{article}
\begin{document}
\pagestyle{plain}
\huge
\title{\bf The bremsstrahlung equation for the spin motion in LHC}
\large
\author{Miroslav Pardy\\[7mm]
Department of Physical Electronics \\
and\\
Laboratory of Plasma physics\\[5mm]
Masaryk University \\
Kotl\'{a}\v{r}sk\'{a} 2, 611 37 Brno, Czech Republic\\
e-mail:pamir@physics.muni.cz}
\date{\today}
\maketitle
\vspace{5mm}

\begin{abstract}
 
The influence of the bremsstrahlung on the spin motion is expressed by the equation which is the analogue and generalization of the Bargmann-Michel-Telegdi equation. The new constant is involved in this equation. This constant can be immediately determined by the experimental measurement 
of the spin motion, or it follows from the classical limit of quantum electrodynamics with radiative corrections.

\end{abstract}

\vspace{3mm}

\vspace{7mm}

\section{Introduction}

The synchrotron radiation evidently influences the motion of
the electron in accelerators. The corresponding equation which describes
the classical motion is so called the Lorentz-Dirac equation, which
differs from the the so called Lorentz equation only by the additional
term which describes the radiative corrections. The equation with the
radiative term is as follows (Landau et al., 1988):

$$m c \frac{du_{\mu}}{ds} = \frac{e}{c}F_{\mu\nu}u^{\nu} +
g_{\mu},\eqno(1)$$
where the radiative term was derived by Landau et al. in the form 
(Landau et al., 1988):

$$g_{\mu} = \frac{2e^{3}}{3mc^{3}}\frac{\partial F_{\mu\nu}}
{\partial x^{\alpha}}u^{\nu}u^{\alpha} - 
\frac{2e^{4}}{3m^{2}c^{5}} F_{\mu\alpha}F^{\beta\alpha}u_{\beta} + 
\frac{2e^{4}}{3m^{2}c^{5}} \left(F_{\alpha\beta}u^{\beta}\right) 
\left(F^{\alpha\gamma}u_{\gamma}\right)u_{\mu}. \eqno(2)$$

It is possible to show that the space components of the 4-vector force $g_{\mu}$ is of the form (Landau et al., 1988) 

$${\bf f} = \frac{2e^{3}}{3mc^{3}}\left(1 - \frac{v^{2}}{c^{2}}\right)^{-1/2}\left\{\left(\frac{\partial}{\partial t} + ({\bf v}\nabla)\right){\bf E}
+ \frac{1}{c}\left[{\bf v}\left(\frac{\partial}{\partial t} + ({\bf v}\nabla)\right){\bf H}\right]\right\} + $$

$$\frac{2e^{4}}{3m^{2}c^{3}}\left\{{\bf E}\times {\bf H} + \frac{1}{c}{\bf H }\times ({\bf H}\times {\bf v}) + \frac{1}{c}{\bf E}({\bf v}{\bf E})\right\} - $$

$$\frac{2e^{4}}{3m^{2}c^{5}\left(1 - \frac{v^{2}}{c^{2}}\right)}{\bf v}\left\{\left({\bf E} + \frac{1}{c}({\bf v}\times{\bf H})\right)^{2} -
\frac{1}{c^{2}}({\bf E}{\bf v})^{2}\right\}.\eqno(3)$$

Bargmann, Michel and Telegdi (Berestetzkii, 1989;) derived so called 
BMT equation for motion of spin in the electromagnetic field,  in the form 

$$\frac{da_{\mu}}{ds} = \alpha F_{\mu\nu}a^{\nu} - \beta u_{\mu}F^{\nu\lambda}u_{\nu}a_{\lambda},\eqno(4)$$
where $a_{\mu}$ is so called axial vector describing the classical
spin and  constants $\alpha$ and $\beta$ was determined after the comparison of the postulated equations with the 
non-relativistic quantum mechanical limit. The result of such comparison is the final form of so called BMT equations:

$$\frac{da_{\mu}}{ds} = 2\mu F_{\mu\nu}a^{\nu} -2\mu'u_{\mu}F^{\nu\lambda}u_{\nu}a_{\lambda},\eqno(5)$$
where $\mu$ is magnetic moment of electron following directly from the Dirac equation and $\mu'$ is anomalous magnetic moment of electron which can be calculated as the radiative correction to the interaction of electron with electromagnetic field and follows from quantum electrodynamics.
 The BMT equation has more earlier origin. The first attempt to describe the spin motion in electromagnetic field using the special theory of relativity was performed by Thomas (1926). However the basic ideas on the spin motion was established by Frenkel (1926). After appearing the Frenkel basic article,
many authors published the articles concerning the spin motion (Ternov et al., 1980; Tomonaga, 1998). The mechanical model of spin was constructed by Uhlenbeck and Goudsmith (1926), or, in the very sophisticated form by Ohanian (1984) and other authors. However, we know at present time that spin of electron is its  physical attribute which follows only from the Dirac equation. Also the Schr\"odinger Zitterbewegung of the Dirac electron as a point-like particle follows from the Dirac equation.   

 It was shown by Rafanelli and Schiller (1964), (Pardy, 1973) 
that the BMT equation
can be derived from the classical limit, i.e. from the WKB solution 
of the Dirac equation with the anomalous magnetic moment. Equation (5) is also the basic equation of the non-dissipative spintronics.

\section{Equation of motion for the spin-vector}

If we introduce the average value of the vector of spin in the rest system by the quantity $\mbox {\boldmath $\zeta$}$,
 then the 4-pseudovector  $a^{\mu}$  is of the from 
$a^{\mu} = (0, \mbox {\boldmath $\zeta$})$. The momentum four-vector of a particle is $p ^{\mu} = (m, 0)$ in the rest system of a particle. Then the equation 

$$a^{\mu}p_{\mu} = 0 \eqno(6)$$
is valid not only in the rest system of a particle but in the arbitrary system as a consequence of the relativistic invariance. The following general formula is also valid in the arbitrary  system

$$a^{\mu}a_{\mu} = - \mbox {\boldmath $\zeta$}^{2}.\eqno(7)$$

The components of the axial 4-vector $a^{\mu}$ in the reference system where particle is moving with the velocity ${\bf v} = {\bf p}/\varepsilon$ can be obtained by application of the Lorentz transformation to the rest system and they are as follows (Berestetzkii et al., 1989):

$$a^{0} = \frac{|{\bf p}|}{m}\mbox {\boldmath $\zeta$}_{\parallel}, \quad {\bf a}_{\perp} = \mbox {\boldmath $\zeta$}_{\perp}, \quad a_{\parallel} = \frac{\varepsilon}{m}\mbox {\boldmath $\zeta$}_{\parallel}, \eqno(8)$$
where suffices $\parallel, \perp$ denote the components of ${\bf a}$, $\mbox {\boldmath $\zeta$}$ parallel and perpendicular to the direction ${\bf p}$. The formulas for the components can be also rewritten in the more compact form as follows (Berestetzkii et al., 1989):

$${\bf a} = {\mbox {\boldmath $\zeta$}} + \frac{{\bf p}({\mbox {\boldmath $\zeta$}}{\bf p})}{m(\varepsilon + m)}, \quad a^{0} = \frac{{\bf a}{\bf p}}{\varepsilon} = \frac{{\mbox {\boldmath $\zeta$}}{\bf p}}{m}, \quad
{\bf a}^{2} = {\mbox {\boldmath $\zeta$}}^{2} + \frac{({\bf p}{\mbox {\boldmath $\zeta$}})^{2}}{m^{2}}.\eqno(9)$$

The equation for the change of polarization can be obtained immediately from the BMT equation in the following form (Berestetzkii et al., 1989):

$$ \frac{d{\bf a}}{dt} = \frac{2\mu m}{\varepsilon}{\bf a}\times{\bf H} + \frac{2\mu m}{\varepsilon}({\bf a}{\bf v}){\bf E} - \frac{2\mu' \varepsilon}{m}{\bf v}({\bf a}{\bf E})\; + $$

$$+ \frac{2\mu'\varepsilon}{m}{\bf v}({\bf v}({\bf a}\times {\bf H})) +  \frac{2\mu'\varepsilon}{m}{\bf v}({\bf a}{\bf v})({\bf v}{\bf E}), \eqno(10)$$
where we used the relativistic relations $c =1$, $ds = dt\sqrt{1 - v^{2}}$ , $\varepsilon = m\sqrt{1 - v^{2}}$ and the following components of the electromagnetic field (Landau et al., 1988):

$$F^{\mu\nu} = \left(\begin{array}{cccc}
0 & -E_{x} & -E_{y} & -E_{z}\\
E_{x} & 0 & -H_{z} & H_{y}\\
E_{y} & H_{z} & 0 & -H_{x}\\
E_{z} & -H_{y} & H_{x} & 0\\
\end{array} \right) \stackrel {d}{=} ({\bf E}, {\bf H}); \quad F_{\mu\nu} = ({-\bf E }, {\bf H}).\eqno(11)$$

Inserting equation ${\bf a}$ from eq. (9) into  eq. (10) and using equations 

$${\bf p} = \varepsilon{\bf v}, \quad \varepsilon^{2} = {\bf p}^{2} + m^{2}, \quad \frac{d{\bf p}}{dt} = e{\bf E} + e({\bf v}\times {\bf H}),\quad
\frac{d\varepsilon}{dt} = e({\bf v}{\bf E}),\eqno(12)$$
we get after long but simple mathematical operations the following equation for the polarization $\mbox {\boldmath $\zeta$}$

$$\frac{d {\mbox {\boldmath $\zeta$}}}{dt} = 
\frac{2\mu m  + 2\mu'(\varepsilon - m)}{\varepsilon}{\mbox {\boldmath $\zeta$}}\times {\bf H} + $$

$$\frac{2\mu' \varepsilon}{\varepsilon + m}({\bf v}{\bf H})({{\bf v}\times \mbox {\boldmath $\zeta$}}) + \frac{2\mu m  + 2\mu' \varepsilon}{\varepsilon + m}{\mbox {\boldmath $\zeta$}}\times ({\bf E}\times {\bf v})\eqno(13)$$

The special interest is concerned not only on the change of the absolute quantity of the polarization, but on the change with regard to the direction of motion represented by the unit vector ${\bf n} = {\bf v}/v$. We write the ploarization in the form:

$$\mbox {\boldmath $\zeta$} = {\bf n}{\zeta}_{\parallel} + {\mbox {\boldmath $\zeta$}_\perp} .\eqno(14)$$

Then using eqs. (12), (13) and (14), we get the following equation for the parallel component of the polarization (Berestetzkii et al., 1989):

$$\frac{d{\zeta}_{\parallel}}{dt} = 2\mu'(\mbox {\boldmath $\zeta$}_{\perp}({\bf H}\times{\bf n})) + \frac{2}{v}\left(\frac{\mu m^{2}}{\varepsilon^{2}} - \mu'\right) (\mbox {\boldmath $\zeta$}_{\perp}{\bf E}).\eqno(15)$$

\section{Spin motion equation with the bremsstrahlung reaction}

It is meaningful to consider the BMT equation with the radiative
corrections to express the influence of the synchrotron radiation on
the motion of spin. To our knowledge such equation, the generalized
BMT equation, was not published
and we here present the conjecture of the form of such equation. 
The equation of the spin motion under the influence of the synchrotron radiation is suggested as an analogue to the BMT construction:

$$\frac{da_{\mu}}{ds} = 2\mu F_{\mu\nu}a^{\nu} -2\mu'u_{\mu}
F^{\nu\lambda}u_{\nu}a_{\lambda} + \Lambda f_{\mu}(axial),\eqno(16)$$
where the term $f_{\mu}(axial)$ is generated as the "axialization" of
the force elaborated from the radiation term $g_{\mu}$. The axialization is the operation which was used by Bargmann, Michel and Telegdi and it consists in the construction of the axial vector from the four-vector force. We see from the right side of the BMT equation how to construct such axial equation. Or, the additional axial 4-vector constructed from the bremsstrahlung force is as following:

$$f_{\mu}(axial) = \Lambda u_{\mu}(g^{\alpha}a_{\alpha}) = \Lambda u_{\mu}[g_{0}a_{0} - {\bf g}\cdot {\bf a}].\eqno(17)$$

So, the generalized BMT equation which involves also the influence of synchrotron radiation on spin motion is as follows: 

$$\frac{da_{\mu}}{ds} = 2\mu F_{\mu\nu}a^{\nu} -2\mu'u_{\mu}
F^{\nu\lambda}u_{\nu}a_{\lambda} + $$

$$\Lambda u_{\mu}\left\{\frac{2e^{3}}{3mc^{3}}\frac{\partial F_{\lambda\nu}}
{\partial x^{\alpha}}u^{\nu}u^{\alpha} - \right. $$

$$\left.\frac{2e^{4}}{3m^{2}c^{5}} F_{\lambda\alpha}F^{\beta\alpha}u_{\beta} + 
\frac{2e^{4}}{3m^{2}c^{5}} \left(F_{\alpha\beta}u^{\beta}\right) 
\left(F^{\alpha\gamma}u_{\gamma}\right)u_{\lambda}\right\}a^{\lambda}. \eqno(18)$$

Using eq. (17), we can write eq. (18) in the form

$$\frac{da_{\mu}}{ds} = 2\mu F_{\mu\nu}a^{\nu} -2\mu'u_{\mu}
F^{\nu\lambda}u_{\nu}a_{\lambda} + \Lambda u_{\mu}[g_{0}a_{0} - {\bf g}\cdot {\bf a}].\eqno(19)$$

The constant $\Lambda$ is new physical constant, which cannot be determined from the classical theory of the spin motion. This constant can be determined immediately  from the spin motion observed experimentally. On the other hand, this constant follows from the classical limit of quantum electrodynamics (QED)  involving radiative corrections. The solution of this problem was not still published.

While the vector component is involved in the equation (3) the zero component must be determined by the extra way. We have:

$$g_{0} = P_{1} + P_{2} + P_{3}, \eqno(20)$$ 
where the terms of eq. (20) follow from eq. (2) in the form (c = 1):

$$P_{1} = \left(\frac{2e^{3}}{3m}\right)\frac{\partial F_{0\nu}}
{\partial x^{\alpha}}u^{\nu}u^{\alpha}, \eqno(21)$$

$$P_{2}  =  \left( - \frac{2e^{4}}{3m^{2}}\right) F_{0\alpha}F^{\beta\alpha}u_{\beta}, \eqno(22)$$

$$P_{3} = \left(\frac{2e^{4}}{3m^{2}}\right) \left(F_{\alpha\beta}u^{\beta}\right) 
\left(F^{\alpha\gamma}u_{\gamma}\right)u_{0} \eqno(23)$$
with

$$u = \left(\frac{1}{\sqrt{1 - v^{2}}}, \frac{\bf v}{\sqrt{1 - v^{2}}}\right). \eqno(24)$$  

After some algebraic operations, we write the set of quantities $P_{1}, P_{2}, P_{3}$ as follows:

$$P_{1} = \left(\frac{2e^{3}}{3m}\right)\frac{1}{1 - v^{2}}\; \times$$

$$\left\{(\partial_{t}{\bf E})\cdot {\bf v}  + (\partial_{x}{\bf E})\cdot {\bf v} v_{x} + 
(\partial_{y}{\bf E})\cdot {\bf v} v_{y}  + (\partial_{z}{\bf E})\cdot {\bf v} v_{z}\right\},\eqno(25)$$

$$P_{2}  =  \left(\frac{2e^{4}}{3m^{2}}\right)\frac{1}{\sqrt{1 - v^{2}}}\left\{E^{2} + ({\bf H}\times{\bf E})\cdot {\bf v}) \right\} \eqno(26)$$

$$P_{3} = \left(\frac{2e^{4}}{3m^{2}}\right) \frac{1}{\left(1 - v^{2}\right)^{3/2}}\left\{ \left({\bf E} + ({\bf v}\times {\bf H})\right)^{2} - 
({\bf E}\cdot {\bf v})^{2}\right\}. \eqno(27)$$

The relation of this equation to the (dissipative) spintronics cannot be a priori excluded.
Such equation will have fundamental meaning for the work of LHC where the synchrotron
radiation influences the spin motion of protons in LHC.

\section{Discussion}

We have considered here the influence of the synchrotron radiation on the spin motion of a charged particle moving in the homogenous magnetic field. It is well known that the synchrotron radiation also influences the trajectory of the charged particle. However we do not consider this influence. It is well known that not only the the synchrotron radiation is produced during the motion of a particle in the magnetic field but also the so called spin light, which is due the spin motion of a particle. We suppose that the influence of the spin light on the spin motion is so small that it is possible to neglect such effect although from the theoretical point of view, it is serious effect to be as the integral part of the theoretical physics. 

The intensity of the synchrotron radiation is, as it is well known,  given by the formula (Ternov, 1994; Bordovitsyn et al., 1995):

$$W_{class.\; synch.\; rad.} = \frac{2}{3}\frac{e^{2}c}{R^{2}}\gamma^{4}; \quad \gamma = \frac{\varepsilon}{m_{0}c^{2}}, \eqno(28)$$
where $R$ is the radius of the circular motion, $\varepsilon$ is the energy of the moving particle.

The intensity of the spin light is expressed by the formula:

$$W_{spin\; light} = \frac{2}{3}\frac{1}{c^{3}}\left(\frac{d^{2}}{dt^{2}}\mbox {\boldmath $\mu$}\right)^{2} = \frac{2}{3}\frac{\mu_{0}^{2}}{c^{3}}\omega_{R}^{4}\zeta_{\perp}^{2}. \eqno(29)$$

After comparison of formula (28) and (29), we see that the the intensity of the spin light is smaller than the intensity of the synchrotron radiation. So, the influence of the spin light on the spin motion can be neglected.

There is the second possibility how to generalize the BMT equation. It consist in axialization of the bremsstrahlung force in the following way:

$$g_{\mu}(axial) =  $$

$$\frac{2e^{3}}{3mc^{3}}\frac{\partial F_{\mu\nu}}
{\partial x^{\alpha}}u^{\nu}a^{\alpha} - 
\frac{2e^{4}}{3m^{2}c^{5}} F_{\mu\alpha}F^{\beta\alpha}a_{\beta} + 
\frac{2e^{4}}{3m^{2}c^{5}} \left(F_{\alpha\beta}u^{\beta}\right) 
\left(F^{\alpha\gamma}u_{\gamma}\right)a_{\mu}. \eqno(30)$$

Then, such force multiplied with the appropriate constant can be add to the original BMT equation. We think that the second conjecture 
which is presented in this article cannot be a priory excluded. 

The verification of the bremsstrahlung equation (16) can be evidently verified by all circular accelerators over the world including the most gigantic LHC which started its activity by 10. 9. 2008.

\vspace{15mm} 
\noindent 
{\bf References}.

\vspace{15mm} 

\noindent
Berestetzkii, V. B., Lifshitz, E. M. and Pitaevskii L. P., (1989). {\it Quantum electrodynamics}, 3rd ed., (Moscow, Nauka). (in Russian).\\[3mm]
Bordovitsyn, V. A., Ternov, I. M. and Bagrov, V. G., (1995). Spin light. Uspekhi fizicheskich nauk {\bf 165}, No. 9, 1083. (in Russian).\\[3mm]
Frenkel, J. I., (1926). Die Elektrodynamik der rotierenden Elektronen, Zs. Physik {\bf 37}, 243. \\[3mm]
Frenkel, J. I.., (1958). {\it Collective scientific works}, II., {\it Scientific articles}, AN SSSR, (in Russian).\\[3mm]
Landau, L. D. and Lifshitz, E. M., (1988). {\it The classical theory of fields}, 7th ed.,
(Moscow, Nauka), (in Russian).\\[3mm]
Ohanian, H. C., (1986). What is spin?, Am. J. Phys. {\bf 54}(6), 500. \\[3mm]
Pardy, M. (1973). Classical motion of spin 1/2 particles with zero anomalous magnetic moment,
Acta Phys. Slovaca {\bf 23}, No. 1, 5. \\[3mm]
Rafanelli, K, and Schiller, R., (1964). Classical motion of spin-1/2 particles, Phys. Rev.
{\bf 135}, No. 1 B, B279. \\[3mm]
Ternov, I. M. (1980). On the contemporary interpretation of the classical theory of the I. A. Frenkel spin, Uspekhi fizicheskih nauk, {}{\bf 132}, 345. (in Russian). \\[3mm]    
Ternov, I. M., (1994). Synchrotron radiation, Uspekhi fizicheskih nauk, {\bf 164}(4), 429. (in Russian).\\[3mm]
Thomas, L. H., (1926). The motion of spinning electron, Nature, {\bf 117}, 514. \\[3mm]
Tomonaga, S.-I., (1997). {\it The story of spin}, (The university of Chicago press, Ltd., London).\\[3mm] 
Uhlenbeck, G. E. and Goudsmit, S. A. (1926). Spinning electrons and the structure of spectra, Nature {\bf 117}, 264.
\end{document}